\documentclass[reprint,twocolumn,prapplied,amsmath,amssymb,superscriptaddress]{revtex4-2}

\usepackage[space]{grffile}	
\usepackage{graphicx}
\graphicspath{
	{./}
}

\usepackage[caption=false]{subfig}
\usepackage{hyperref}
\usepackage{cleveref}
\usepackage{float}
\usepackage{physics}
\newcommand{\comsol}{COMSOL Multiphysics\textsuperscript{\textregistered}}
\newcommand{\hc}{\mathrm{h.c}}

\usepackage{multirow} 

\usepackage[todonotes={textsize=small},defaultcolor=cyan,authormarkuptext=name,commentmarkup=todo,final]{changes}
\definechangesauthor[name=Alan]{al}
\definechangesauthor{0}
\definechangesauthor{1}
\definechangesauthor{2}
\definechangesauthor{3}
\definechangesauthor{4}
\definechangesauthor{5}
\definechangesauthor[name={}]{typo}
\let\oldcite\cite
\renewcommand{\cite}[1]{\mbox{\oldcite{#1}}}

\setdeletedmarkup{}

\newcommand{\phantomsubfloat}[1]{
	{
		\captionsetup[subfigure]{labelformat=empty}
		\subfloat[][]{#1}
	}%
}

\begin{document}
    \title{Engineering synthetic gauge fields through the coupling phases in cavity magnonics}
    
    \author{Alan Gardin}
    \email{alan.gardin@adelaide.edu.au}
    \affiliation{%
    	School of Physics, The University of Adelaide, Adelaide SA 5005, Australia
    }%
    \affiliation{%
    	IMT Atlantique, Lab-STICC, UMR CNRS 6285, F-29238 Brest, France
    }%
	\author{Guillaume Bourcin}
	\affiliation{%
		IMT Atlantique, Lab-STICC, UMR CNRS 6285, F-29238 Brest, France
	}%
    \author{Jeremy Bourhill}
    \affiliation{%
    	ARC Centre of Excellence for Engineered Quantum Systems and ARC Centre of Excellence for Dark Matter Particle Physics, \\
    	Department of Physics, University of Western Australia,	35 Stirling Highway, Crawley, Western Australia 6009, Australia
    }%
    \author{Vincent Vlaminck}
    \affiliation{%
    	IMT Atlantique, Lab-STICC, UMR CNRS 6285, F-29238 Brest, France
    }%
    \author{Christian Person}
    \affiliation{%
    	IMT Atlantique, Lab-STICC, UMR CNRS 6285, F-29238 Brest, France
    }%
    \author{Christophe Fumeaux}
    \affiliation{%
    	School of Electrical Engineering and Computer Science, The University of Queensland, Brisbane QLD 4072, Australia
    }%
    \author{Giuseppe C. Tettamanzi}
    \affiliation{%
    	School of Physics, The University of Adelaide, Adelaide SA 5005, Australia
    }%
    \affiliation{%
    	School of Chemical Engineering and Advanced Materials, The University of Adelaide, Adelaide SA 5005, Australia
    }%
    \author{Vincent Castel}
    \affiliation{%
    	IMT Atlantique, Lab-STICC, UMR CNRS 6285, F-29238 Brest, France
    }%

    \begin{abstract}
    	Cavity magnonics, which studies the interaction of light with magnetic systems in a cavity, is a promising platform for quantum transducers\replaced[id=1]{ and quantum memories}{, quantum memories, and devices with non-reciprocal behaviour}. At microwave frequencies, the coupling between a cavity photon and a magnon, the quasi-particle of a spin wave excitation, is a consequence of the Zeeman interaction between the cavity's magnetic field and the magnet's macroscopic spin. For each photon/magnon interaction, a coupling phase factor exists, but is often neglected in simple systems.     
    	However, in ``loop-coupled'' systems, where there are at least as many couplings as modes, the coupling phases become relevant for the physics and lead to synthetic gauge fields.
    	We present experimental evidence of the existence of such coupling phases by considering two spheres made of Yttrium-Iron-Garnet and two different re-entrant cavities. We predict numerically the values of the coupling phases, and we find good agreement between theory and the experimental data. These results show that in cavity magnonics, one can engineer synthetic gauge fields, which can be useful for cavity-mediated coupling\replaced[id=1]{ and engineering dark mode physics}{, dark mode memories, and building nonreciprocal devices}.
    \end{abstract}
    \maketitle

    \section{Introduction}
    Magnons are the quasi-particles associated with the elementary excitation of a spin wave in a magnetically ordered material, such as the ferrimagnet Yttrium-Iron-Garnet (YIG). The field of cavity magnonics, or spin cavitronics, aims to use the interaction of photons in a cavity with magnons for both classical (e.g. radio-frequency circulators or isolators, and spintronics 
    \cite{2001WolfAwschalom,2015Chumak,2017Csaba,2017Chumak,2019Chumak,2021Barman,2020Mahmoud,2017Bhatti,2022RoadmapSpinWaveComputingConcepts,2020Grollier}) and quantum technologies \cite{2021HarderHu,2021Rameshti,2022Yuan}. Magnons are notably promising for quantum transduction \cite{2019LachanceQuirion}, due to their capability to couple to phonons \cite{2016Zhang}, optical and microwave photons, and superconducting qubits \cite{2015Tabuchi,2016Tabuchi,2017LachanceQuirion,2020LachanceQuirion}. The coupling with microwave photons has been particularly fruitful, with demonstrations of coherent coupling \cite{2021Crescini}, indirect coupling \cite{2016Lambert,2016Hyde}, ultrastrong coherent coupling \cite{2014Goryachev,2016Kostylev,2021Golovchanskiy,2021Golovchanskiya,2022Bourcin}, dissipative coupling \cite{2019Yu,2020Zhao,2018Harder,2019Bhoi,2019Yang,2019Yao,2019Yaoa,2021Lu,2019Wang}, the tuning between level repulsion and attraction \cite{2019Boventer,2020Boventer}, a dark mode memory \cite{2015Zhang}, and non-reciprocal effects \cite{2019Wang,2022Li,2020Zhu,2022Kong}.
    
    Physically, all of these results involve the photon/magnon coupling, which originates from the Zeeman interaction between the magnetic dipole moment of the magnet, and the magnetic field. This interaction is characterised by a coupling strength and a coupling phase, which can both be chosen real positive numbers. \added[id=0]{As recently shown by \cite{2023Gardin},} \replaced[id=0]{in most simple systems}{In most simple systems,} the coupling phases can be ignored since they do not affect the physics. However, in ``loop-coupled'' systems \cite{2022Kong,2019Zheng,2023Huang,2010Koch,2017Bernier,2017Fang,2021Chen}, where there are at least as many couplings as modes, neglecting the coupling phases can lead to an erroneous description of the physics.
    \deleted[id=1]{Indeed in these systems, the various coupling phases (each due to a photon/magnon interaction) combine to form synthetic $U(1)$ gauge fields characterised by one gauge-invariant phase $\theta$ per loop. These synthetic gauge fields can be used to balance coherent and dissipative couplings \cite{2023Biehs,2022Clerk,2015Metelmann} to engineer non-reciprocal interactions, with proposals in circuit QED \cite{2015Metelmann,2010Koch,2015Sliwa} and cavity optomechanics \cite{2017Bernier,2017Fang,2021Chen}.}\replaced[id=4]{Notably, in the context of cavity magnonics, it was theoretically shown that the coupling phases}{In cavity magnonics, it was theoretically shown that a synthetic gauge field} can steer the system towards either dark mode physics (useful for quantum memories \cite{2015Zhang}), or the enhancement of cavity-mediated coupling by taking advantage of several cavity modes \cite{2023Gardin}, useful for quantum information transduction.

    \replaced[id=0]{However, so far no experimental evidence of the coupling phases have been reported, despite their striking impact on cavity magnonics applications.}{Recently, the influence of the coupling phases was theoretically analysed in a loop-coupled system \cite{2023Gardin} consisting of two magnon modes coupling to two cavity modes, however without experimental verification.}
    To bridge this gap, we propose two re-entrant three-post cavities, in which multiple cavity modes simultaneously couple to two magnon modes, creating loop-coupled systems. We experimentally measure a first cavity in which the physics is characterised by a single physical phase $\theta = \pi$, which makes it fall within the theoretical analysis of the previous work \cite{2023Gardin}. In contrast, in the second cavity, the physics is determined by two physical phases $\theta_1=0$ and $\theta_2=\pi$. In both cavities, we predict theoretically the values of the physical phases, and find good agreement with our experiments. 
	
	In \cref{sec:coupling-phase}, we introduce the coupling phases, and how one can reduce them to  the so-called \emph{physical phases} characterising the physics using unitary transformations. \added[id=2]{Importantly, we interpret these physical phases as parametrising a synthetic $U(1)$ gauge field (or synthetic magnetic field), allowing to connect our work to the wider literature.} Next, in \cref{sec:single-phase}, we introduce a cavity with a physical phase $\theta=\pi$, which is supported by both theory and experiment. In \cref{sec:two-phases}, we perform a similar analysis, but this time on a cavity leading to physical phases $\theta_1=\pi$ and $\theta_2 = 0$. We conclude in \cref{sec:conclusion}.

	\section{\label{sec:coupling-phase}Introduction to the magnon/photon coupling phase}
	We adopt a second-quantised formalism for the description of the cavity (annihilation operator $c_k$) and magnon modes (annihilation operator $m_l$ after performing the Holstein-Primakoff transformation \cite{1940Holstein} on the macrospin operator \cite{2019FlowerGoryachevBourhillTobar}). The ferromagnetic resonance frequency of the magnons is tuned by a static applied magnetic field $\vb{H}_0 = H_0 \vu{z}$. The coupling between the cavity mode $c_k$ and the magnon mode $m_l$ is characterised by a coupling strength $g_{kl}/2\pi$ and a coupling phase $\varphi_{kl}$ defined as \cite{2019FlowerGoryachevBourhillTobar,2023Gardin}
	\begin{gather}
		\label{eq:coupling-strength}
		g_{kl}/2\pi = \eta_{kl} \sqrt{\omega_k} \frac{\gamma}{4\pi} \sqrt{
			\frac{\mu}{g_L\mu_B} \hbar \mu_0 n_s
		},\\
		\label{eq:coupling-phase}
		\varphi_{kl} = \arg \qty{
			\int_{V_m} \dd[3]{r} \vb{h}_k(\vb{r}) \cdot \vu{x} 
			+ i \int_{V_m} \dd[3]{r} \vb{h}_k(\vb{r}) \cdot \vu{y}
		},
	\end{gather}
	where $\gamma/2\pi = 28$ GHz/T is the gyromagnetic ratio, $\mu = 5 \mu_B$ is the magnetic moment of a unit cell of YIG, {$\mu_B$ is the Bohr magneton, $\mu_0$ is the magnetic permeability of vacuum}, $g_L=2$ is the Landé $g$-factor, $n_s = 4.22 \times 10^{23}$ m\textsuperscript{-3} is the spin density of YIG \cite{2020BourhillCastel}, $\vb{h}_k$ is the magnetic field vector of the cavity mode $k$, $V_m$ is the volume of the YIG spheres, and
	\begin{equation}
		\label{eq:filling-factor}
		\eta_{kl} = \sqrt{
			\frac{
				\qty(\int_{V_m} \dd[3]{r} \vb{h}_k(\vb{r}) \cdot \vu{x})^2
				+ \qty(\int_{V_m} \dd[3]{r} \vb{h}_k(\vb{r}) \cdot \vu{y})^2
			}{
				V_m\int_{V_c} \dd[3]{r} \abs{\vb{h}_k(\vb{r})}^2
			}
		}
	\end{equation}
	is the filling-factor, with $V_c$ the volume of the cavity. 
	
	The interaction Hamiltonian between the cavity mode $c_k$ and the magnon mode $m_l$ reads \cite{2023Gardin}
	\begin{equation}
		\label{eq:interaction-hamiltonian}
		H_I = \hbar g_{kl} \qty(e^{i \varphi_{kl}}c_km_l^\dagger + e^{-i \varphi_{kl}}c_k^\dagger m_l).
	\end{equation}
	where we used the rotating wave approximation, which is valid provided $\frac{g_{kl}}{\omega_k} < 10\%$ \cite{2019FriskKockum,2020LeBoite}. Note that the interaction Hamiltonian \cref{eq:interaction-hamiltonian} is Hermitian, and results \replaced[id=typo,comment={}]{from}{for} the coherent coupling between photons and magnons. This is in stark contrast with dissipative couplings \cite{2021HarderHu,2021Rameshti}, when the interaction is non-Hermitian and results from an indirect coupling \cite{2015Metelmann} of the magnon and photons through a strongly dissipative mode \cite{2019Yu,2020Zhao}, or travelling wave reservoir \cite{2019Wang,2018Harder,2019Bhoi,2019Yang,2019Yao,2019Yaoa,2021Lu}.
	
	Formally, recall that the unitary transformation $U = e^{i \varphi c^\dagger c}$ transforms the operator $c$ as $c \mapsto e^{-i \varphi}c$, and that unitary transformations amount to a change of basis. Thus, the transformations $U_{c_k} = e^{i \varphi_{kl} c_k^\dagger c_k}$ and $U_{m_l} = e^{-i \varphi_{kl} m_l^\dagger m_l}$ can both remove the coupling phase from \cref{eq:interaction-hamiltonian}. Physically, this means that there exists a basis in which the coupling phase vanishes, so we might as well set it to zero to simplify the study of the physics. In other words, this amounts to changing the reference phase.
	
	The situation is more complicated when one considers several simultaneous couplings. Naturally, any isolated bosonic mode has a local phase degree of freedom (corresponding to a $U(1)$ symmetry), since e.g. the mapping $c \mapsto e^{-i \varphi}c$ does not change its Hamiltonian. However, couplings between modes constrain these local phase choices, which can break this symmetry. For instance, the coupling of \cref{eq:interaction-hamiltonian} ``locks'' the local phase choices of $c_k$ and $m_l$ together, but we are still free to rotate either $c_k$ (using $U_{c_k}$) or $m_l$ (using $U_{m_l}$) to remove the coupling phase. In other words, we have two phase degrees of freedom, but only one constraint between them. However, when the number of couplings (constraints) is greater or equal to the number of modes (degrees of freedom), then we cannot remove all the coupling phases, and they become relevant for the physics.
	\onecolumngrid
	
	\begin{figure}[H]
		\centering
		\includegraphics[width=0.9\linewidth]{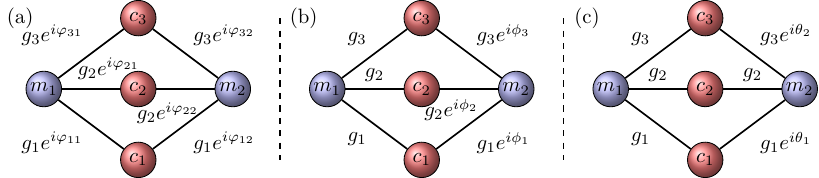}
		\phantomsubfloat{\label{fig:phase-reduction:a}}
		\phantomsubfloat{\label{fig:phase-reduction:b}}
		\phantomsubfloat{\label{fig:phase-reduction:c}}
		\vspace{-1em}
		\caption{Simplification of the physics of a system composed of two magnon modes $m_1$ and $m_2$ (in blue) coupling equally and simultaneously to three cavity modes $c_1,c_2$ and $c_3$ (in red). (a) The system where all the coupling phases remain. (b) After unitarily transforming the cavity modes using \cref{eq:rotation-ci}, the physics is determined only by three phases $\phi_k = \varphi_{k2}-\varphi_{k1}$, accounting for the difference in coupling for each cavity mode $k \in \qty{1, 2, 3}$. (c) After rotating $m_2$ using \cref{eq:rotation-m2}, we are left with two physical phases $\theta_1 = \phi_1 - \phi_2$ and $\theta_2 = \phi_3 -\phi_2$.}
		\label{fig:phase-reduction}
	\end{figure}
	\vspace{1em}

	\begin{figure}[H]
		\centering	
		\includegraphics[width=\linewidth]{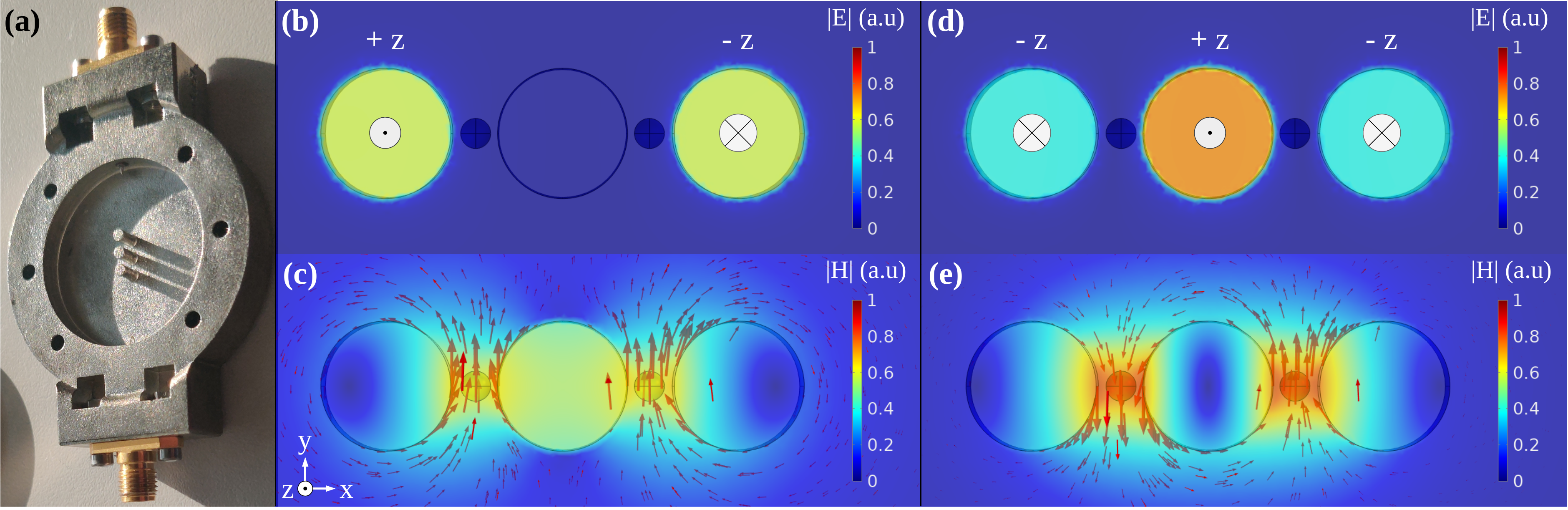}
		\phantomsubfloat{\label{fig:cavity-pi}}
		\phantomsubfloat{\label{fig:cavity-pi-eigenmode-1-electric}}	
		\phantomsubfloat{\label{fig:cavity-pi-eigenmode-1-magnetic}} 
		\phantomsubfloat{\label{fig:cavity-pi-eigenmode-2-electric}}
		\phantomsubfloat{\label{fig:cavity-pi-eigenmode-2-magnetic}}
		\vspace{-2em}
		\caption{(a) The three-posts re-entrant cavity $\pi$ used in the experiment. The lid of the cavity is not shown. (b-c) Electric and magnetic field norm simulated using \comsol\ of the first eigenmode $c_1$ (frequency $\omega_{c,1}/2\pi=4.524$ GHz) of cavity $\pi$. (d-e) Electric and magnetic field distribution simulated using COMSOL of the second eigenmode $c_2$ (frequency $\omega_{c,2}/2\pi=6.378$ GHz) of cavity $\pi$. For each mode, the magnetic field is maximal between the posts, where the YIG spheres are placed, while the electric field is concentrated between the top of the posts and the lid of the cavity. The vectors are only shown for the magnetic field}
		\label{fig:cavity-pi:all}
	\end{figure}
	\twocolumngrid\noindent

	To illustrate this concept, let us consider the example of two magnon modes simultaneously coupling to three cavity modes, as illustrated in \cref{fig:phase-reduction}. We can eliminate the coupling phases between $m_1$ and the cavity modes in \cref{fig:phase-reduction:a} using 
	\begin{equation}
		\label{eq:rotation-ci}
		U_{c_k} = e^{i \varphi_{k1}c_k^\dagger c_k}, \quad k \in \qty{1, 2, 3},
	\end{equation}
	and we obtain \cref{fig:phase-reduction:b}.	Next, we rotate $m_2$ with
	\begin{equation}
		\label{eq:rotation-m2}
		U_{c_k} = e^{i \phi_{2}m_2^\dagger m_2},
	\end{equation}
	and finally obtain two \emph{physical phases} $\theta_1$ and $\theta_2$ characterising the physics (see \cref{fig:phase-reduction:c}).

	Therefore, in the system of \cref{fig:phase-reduction}, we have six couplings and five modes, leading to two physical phases. If the two magnons couple to only two cavity modes (e.g. by setting $g_3=0$), we have as many constraints as degrees of freedom, and the 
	system is characterised by a single physical phase.
	
	\added[id=2]{To conclude this section, we would like to mention that a physical phase effectively parametrises an artificial (or synthetic) $U(1)$ gauge field. Indeed, \cref{eq:interaction-hamiltonian} can be interpreted as the interaction Hamiltonian of a charged particle on a lattice, in the presence of a static magnetic field, see \cite{2010Koch,2022Clerk} for instance. These artificial gauge fields have been proposed for several physical systems, such as neutral atoms \cite{2009Lin,2011Dalibard}, circuit QED \cite{2015Metelmann,2010Koch,2015Sliwa} and cavity optomechanics \cite{2017Bernier,2017Fang,2021Chen}. Electrodynamics is naturally a $U(1)$ gauge theory, and therefore, by analogy, an artificial $U(1)$ gauge field leads to synthetic electromagnetism. Notably, an artificial magnetic field can be used to break time-reversal symmetry in systems where the presence of a ``natural'' magnetic field is problematic, such as in circuit QED \cite{2010Koch}. In turn synthetic gauge fields can be used to create non-reciprocal behaviours by balancing coherent and dissipative couplings through reservoir engineering \cite{2023Biehs,2022Clerk,2015Metelmann}, with proposals for circuit QED \cite{2015Metelmann,2010Koch,2015Sliwa} and cavity optomechanics \cite{2017Bernier,2017Fang,2021Chen}. Framing the effect of the coupling phases in cavity magnonics in terms of synthetic gauge field allows to make a connection with all the aforementioned works. However, it is worth noting that the associated applications may not be so relevant for cavity magnonics, because a static magnetic field is already present. Still, as shown by \cite{2023Gardin}, the coupling phases, or the associated synthetic gauge field, can lead to distinct physics in cavity magnonics. Therefore, their study remains of practical interest.}

	\section{\label{sec:single-phase}A single physical phase}
	\paragraph{Cavity design}
	The first cavity we consider, called cavity $\pi$, is pictured in \cref{fig:cavity-pi}, and contains three posts. It is based on the design of a two-post re-entrant cavity from ref \cite{2014Goryachev,2016Kostylev}, and possesses similar eigenmodes. 
	The distribution of the magnetic field of the two cavity eigenmodes of interest was simulated using \comsol\ (see \cref{fig:cavity-pi-eigenmode-1-magnetic,fig:cavity-pi-eigenmode-2-magnetic}). The magnetic field of these two cavity eigenmodes are concentrated between the posts, where we place two identical YIG spheres of diameter 470 $\mu$m. Importantly, the magnetic field of the eigenmodes is either in the same or opposite directions at the sphere locations, which leads to different coupling phases. This difference can also be understood by considering the strength and direction of the electric field, which is concentrated between the top of the posts and the lid (see \cref{fig:cavity-pi-eigenmode-1-electric,fig:cavity-pi-eigenmode-2-electric}). For instance, the electric field of the first cavity eigenmode (see \cref{fig:cavity-pi-eigenmode-1-electric}) is localised only on the left and right posts, but with opposing directions (which by the right-hand rule, agrees 
	\onecolumngrid
	
	\begin{figure}[H]
		\centering
		\includegraphics[width=\linewidth]{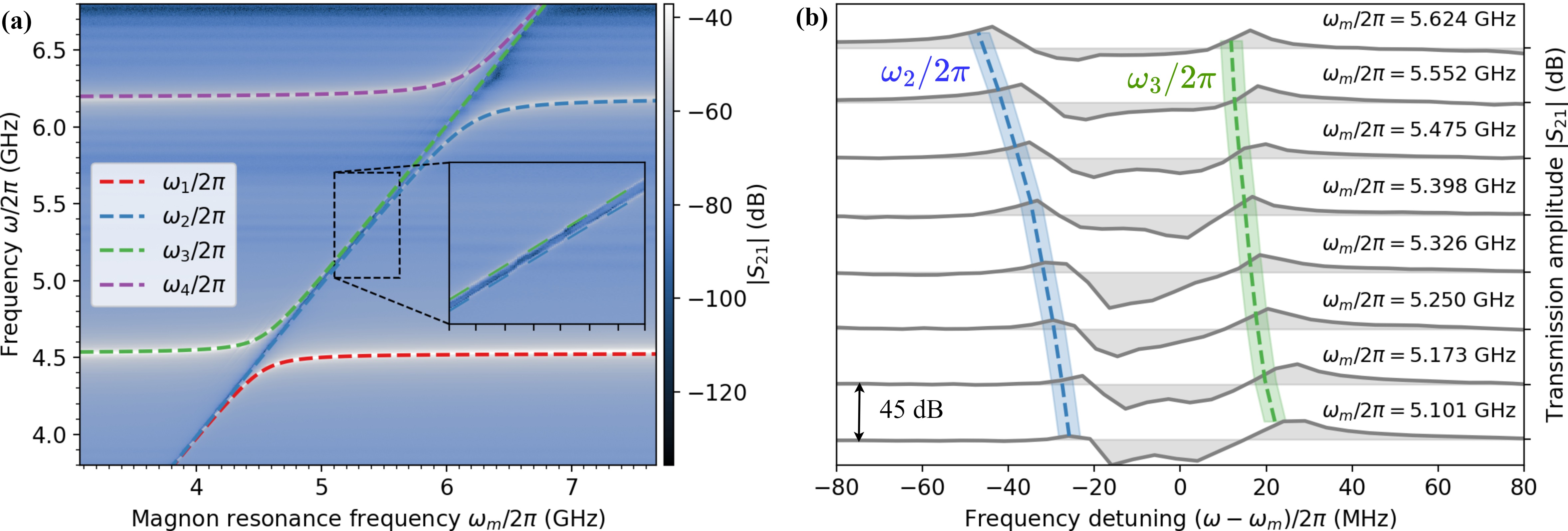}
		\vspace{-2em}
		\phantomsubfloat{\label{fig:cavity-pi:measurement:map}}
		\phantomsubfloat{\label{fig:cavity-pi:measurement:cuts}}
		\caption{(a) Experimental transmission amplitude $\abs{S_{21}}$ of the cavity $\pi$, along with the theoretical spectrum in dashed lines. The fit parameters are  $\omega_{c,1}/2\pi = 4.527$ GHz, $\omega_{c,2}/2\pi=6.19$ GHz , $g_1/2\pi=81$ MHz, $g_2/2\pi=120$ MHz, and $\theta_1=\pi$. Inset: zoom for $5.1 \leq \omega_m/2\pi \leq 5.62$ GHz, and $5.02 \leq \omega/2\pi \leq 5.7$ GHz, showing that the hybridised photon-magnon polariton frequencies $\omega_2/2\pi$ and $\omega_3/2\pi$ do not cross. The ticks indicate the $\omega_m/2\pi$ cuts plotted in (b). (b) Line cuts for different values of $\omega_m/2\pi$ of the transmission amplitude $\abs{S_{21}}$ with the spectrum in dashed lines. The width of $\omega_/2\pi$ and $\omega_3/2\pi$ show the uncertainty $\pm 2.5$ MHz in frequency. For each line cut, the average value of the transmission is used to define the horizontal line from which the colouring begins: this allows to distinguish between resonances and anti-resonances. All the transmission line cuts are separated by 45 dB. The legend for the spectrum is common to (a) and (b).}
		\label{fig:cavity-pi:measurement}
	\end{figure}
	\twocolumngrid\noindent
	with the circulation of the magnetic field shown in \cref{fig:cavity-pi-eigenmode-1-magnetic}). Conversely, for the second cavity eigenmode, the electric field is non-zero on top of each posts, albeit with different intensities and orientations, and thus leads to the magnetic field of \cref{fig:cavity-pi-eigenmode-2-magnetic}.

	\paragraph{Prediction of the physical phase}
	The numerical evaluation of the coupling strengths and coupling phases (given by \cref{eq:coupling-strength,eq:coupling-phase}) can be performed using the eigenmode analysis of \comsol, the results of which are summarised in \cref{tab:cavity-pi-comsol-theory}. Recall that the first index corresponds to the cavity mode, while the second is for the magnon mode. For the magnon modes, $l=1$ ($l=2$) corresponds to the left (right) YIG sphere. We first note that the coupling phases indeed follow the direction of the two cavity eigenmodes's magnetic field shown in \cref{fig:cavity-pi:all}. Furthermore, we observe that both YIG spheres couple with the same coupling strength to each cavity mode, as can be expected from the symmetry of the problem. Hence from now on we set $g_{11} = g_{12} = g_1$ and $g_{21} = g_{22} = g_2$.
	Assuming both YIG spheres to be perfectly identical, the Hamiltonian of the system is 
	$H_\pi = H_\text{free} + H_I$ with
	\begin{equation}
		\label{eq:hamiltonian-free}
		H_\text{free} = \sum_{k=1}^2 \hbar \omega_{c,k} c_k^\dagger c_k + \sum_{l=1}^2 \hbar \omega_m m_l^\dagger m_l
	\end{equation}
	where $\omega_m= \gamma H_0$ is the ferromagnetic resonance frequency of the lowest-order magnon modes (Kittel mode) and
	\begin{equation}
		\begin{aligned}
			H_I &= 
			\;\hbar g_1 e^{-i\frac \pi 2} c_1m_1^\dagger + \hbar g_1 e^{-i\frac \pi 2} c_1 m_2^\dagger \\
			&\quad + \hbar g_2e^{i\frac \pi 2} c_2m_1^\dagger + \hbar g_2 e^{-i\frac \pi 2} c_2m_2^\dagger
			+ \hc
		\end{aligned}
	\end{equation}
	with $\hc$ indicating the hermitian conjugate terms. This system falls within the example of \cref{fig:phase-reduction} by setting $g_3=0$, from which we deduce that after the unitary transformations, the interaction Hamiltonian can be written as
	\begin{equation}
		\label{eq:cavity-pi:interaction-simplified}
		\begin{aligned}
			H_I &= 
			\hbar g_1  c_1m_1^\dagger + \hbar g_1 e^{i \theta_1} c_1 m_2^\dagger \\
			&\quad \hbar g_2 c_2m_1^\dagger + \hbar g_2 c_2m_2^\dagger
			+ \hc,
		\end{aligned}
	\end{equation}
	with $\theta_1 = \varphi_{12}-\varphi_{11}-\qty(\varphi_{22}-\varphi_{21}) =\pi$, hence the name ``cavity $\pi$'' for the cavity.

	\begin{table}[t]
		\caption{Numerical calculation of the coupling strengths $g_{kl}$ and coupling phases $\varphi_{kl}$ of cavity $\pi$ using COMSOL.}
		\centering
		\begin{tabular}{lc|lc}
			\hline
			\multicolumn{2}{c}{Coupling strength (MHz)} & \multicolumn{2}{|c}{Coupling phase (radians)} \\
			\hline
			$g_{11}/2\pi$ & 139 & $\varphi_{11}$ & $-\frac{\pi}{2}$\\
			$g_{12}/2\pi$ & 139 & $\varphi_{12}$ & $-\frac{\pi}{2}$\\
			$g_{21}/2\pi$ & 207 & $\varphi_{21}$ & $+\frac{\pi}{2}$\\
			$g_{22}/2\pi$ & 207 & $\varphi_{22}$ & $-\frac{\pi}{2}$\\
			\hline
		\end{tabular}
		\label{tab:cavity-pi-comsol-theory}
	\end{table}
	
	The values of the coupling strengths and of the physical phase can be verified by the transmission amplitude $\abs{S_{21}}$ through the cavity. Hence in \cref{app:simulations} we performed a frequency domain simulation using COMSOL to numerically compute $\abs{S_{21}}$ as a function of the magnon resonance frequency $\omega_m/2\pi$, and we obtained good agreement with the eigenmode results of \cref{tab:cavity-pi-comsol-theory}.

	\paragraph{Experimental results}
	The three-post cavity $\pi$ design was 3D-printed and then metallised. Using a vector network analyser, we measured the transmission through the cavity when the static magnetic field $\vb{H}_0$ is swept to tune $\omega_m$, and we obtained the data in \cref{fig:cavity-pi:measurement}. The hybridised photon-magnon polariton frequencies are predicted by the spectrum of $H_\pi = H_\text{free} + H_I$, with $H_\text{free}$ given by \cref{eq:hamiltonian-free} and $H_I$ given by \cref{eq:cavity-pi:interaction-simplified}, is plotted on top of the colour map with $\theta_1=\pi$ as suggested by \cref{tab:cavity-pi-comsol-theory}, but with adjusted parameters $g_1/2\pi=81$ MHz and $g_2/2\pi=120$ MHz instead. We attribute the difference between theory and experiment for the coupling strengths to the imperfection of the 3D-printing and metallisation processes. In particular, the posts of cavity $\pi$ are not completely cylindrical which changes the focusing of the magnetic field between the two posts. 
	
	Nevertheless, we note that while the observed coupling strengths are lower than the theoretical values, the prediction for $\theta_1$ remains valid. Indeed, as shown in \cite{2023Gardin}, for $\theta_1 =0$ the frequencies $\omega_2/2\pi$ and $\omega_3/2\pi$ cross, while for $\theta_1=\pi$ they do not. \Cref{fig:cavity-pi:measurement:cuts} and the inset of \cref{fig:cavity-pi:measurement:map} confirm the existence of a frequency gap between $\omega_2/2\pi$ and $\omega_3/2\pi$, which is the signature of the indirect coupling between both magnons, simultaneously mediated by the two cavity modes \cite{2023Gardin}. In other words, this spectral feature demonstrates the coherent coupling between two spatially distant YIG spheres mediated by the two cavity modes, despite the relatively high frequency detuning. Note that if we had $\theta_1 = 0$ instead, the indirect magnon-magnon coupling induced by each cavity mode would interfere destructively. 
	
	The coherent coupling of a cavity mode to two YIG spheres leads to an enhancement of $\sqrt 2$ of the frequency gap at the anticrossings, i.e. the gap is given by $2\sqrt 2 g$ instead of $2g$ \cite{2023Gardin,2022Nair}, which we verified in \cref{app:single-yig-sphere} by comparing with measurements with a single YIG sphere. \added[id=3]{Additionally, by examining the linewidths, we measured $\kappa_1/2\pi=15 \pm 2.5$ MHz and $\kappa_2/2\pi = 22.5 \pm 2.5$ MHz for the linewidths of the cavity modes, $\kappa_m/2\pi=5 \pm 2.5$ MHz for the YIG spheres.}

	\section{\label{sec:two-phases}Analysis of two physical phases}
	
\begin{figure}[t]
		\centering
		\includegraphics[width=\linewidth]{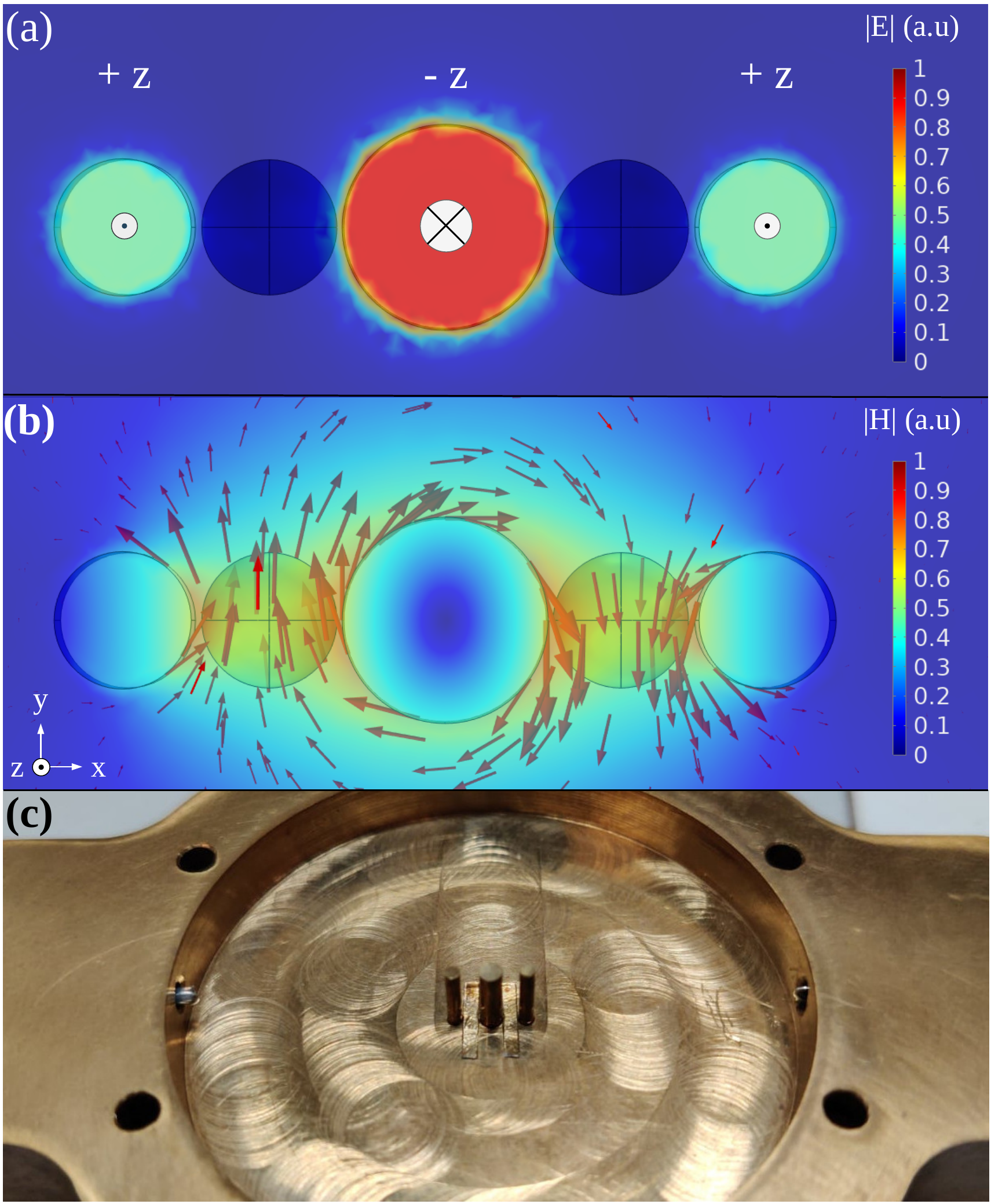}
		\phantomsubfloat{\label{fig:cavity-pi0-electric}} 
		\phantomsubfloat{\label{fig:cavity-pi0-magnetic}}
		\phantomsubfloat{\label{fig:cavity-pi0}}
		\vspace{-3em}
		\caption{Electric (a) and magnetic (b) field distribution simulated using \comsol\ of the second eigenmode $c_2$ of cavity $\pi 0$. The two eigenmodes of cavity $\pi$ still exist but have different frequencies. (c) The three-posts re-entrant cavity $\pi 0$. Contrary to cavity $\pi$, the three posts do not have the same radius, leading to the creation of an additional cavity mode shown in (a) and (b).}
		\label{fig:cavity-pi0-eigenmodes}
	\end{figure}

	\paragraph{Cavity $\pi0$ design} Using cavity $\pi$, we showed the existence of a single physical phase $\theta_1=\pi$. In this section, we examine the possibility of obtaining a different value for the physical phase. In the previous cavity, the origin of the difference in circulation of the magnetic field between the posts can be understood by examining the distribution of the electric field. Following this principle, we engineered another 3-posts re-entrant cavity to obtain two different values $\theta=0$ and $\theta = \pi$ for the physical phases in a unique cavity. This is achieved by making the radius of the centre post (0.75 mm) larger than that of the other two posts (0.5 mm), as shown in \cref{fig:cavity-pi0-eigenmodes}. As a result of this difference in the radii, the electric field intensity is stronger on the centre post compared to the other two (see \cref{fig:cavity-pi0-electric}). This leads to the appearance of another cavity eigenmode of frequency $\omega_{c,2}/2\pi=7.525$ GHz in which the magnetic field distribution solely rotates around the centre post as shown in \cref{fig:cavity-pi0-magnetic}. Interestingly, the other two cavity eigenmodes observed for cavity $\pi$ remain, but their frequencies is modified as $\omega_{c,1}/2\pi \mapsto \omega_{c,1}/2\pi = 6.544$ GHz and $\omega_{c,2}/2\pi \mapsto \omega_{c,3}/2\pi = 8.567$ GHz.  

	\paragraph{Prediction of the physical phases}	
	\begin{table}[b]
		\caption{Numerical calculation of the coupling strengths $g_{kl}$ and coupling phases $\varphi_{kl}$ of cavity $\pi 0$ using COMSOL.}
		\centering
		\begin{tabular}{lc|lc}
			\hline
			\multicolumn{2}{c}{Coupling strength (MHz)} & \multicolumn{2}{|c}{Coupling phase (radians)} \\
			\hline
			$g_{11}/2\pi$ & 130 & $\varphi_{11}$ & $-\frac{\pi}{2}$\\
			$g_{12}/2\pi$ & 127 & $\varphi_{12}$ & $-\frac{\pi}{2}$\\
			$g_{21}/2\pi$ & 148 & $\varphi_{21}$ & $+\frac{\pi}{2}$\\
			$g_{22}/2\pi$ & 150 & $\varphi_{22}$ & $-\frac{\pi}{2}$\\
			$g_{31}/2\pi$ & 103 & $\varphi_{31}$ & $+\frac{\pi}{2}$\\
			$g_{32}/2\pi$ & 104 & $\varphi_{32}$ & $-\frac{\pi}{2}$\\
			\hline
		\end{tabular}
		\label{tab:cavity-pi0}
	\end{table}
	
	The resulting system consists of three cavity eigenmodes, each coupling to the two YIG spheres. The coupling strengths and coupling phases are computed using COMSOL, with the results listed in \cref{tab:cavity-pi0}. Once again, given the symmetry of the problem, both YIG spheres couple with almost equal coupling strength to each cavity eigenmode, so we set $g_{11} = g_{12} = g_1 = 130$ MHz, $g_{21} = g_{22} = g_2 = 150$ MHz and $g_{31} = g_{32} = g_3 = 104$ MHz. The system thus~%
	\vspace{1em}
	\onecolumngrid
	
	\begin{figure}[H]
		\centering
		\includegraphics[width=\linewidth]{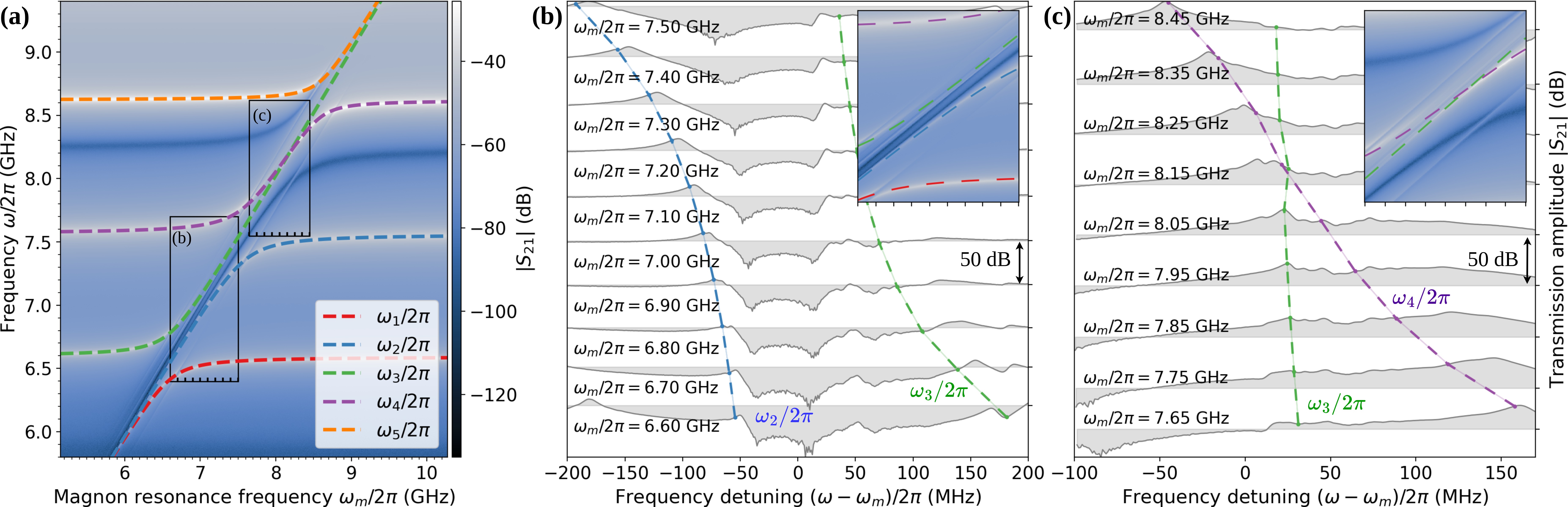}
		\phantomsubfloat{\label{fig:cavity-pi0:measurement:map}}
		\phantomsubfloat{\label{fig:cavity-pi0:measurement:cuts-1}}
		\phantomsubfloat{\label{fig:cavity-pi0:measurement:cuts-2}}
		\vspace{-3em}
		\caption{(a) Experimental transmission amplitude ($\abs{S_{21}}$) of cavity $\pi 0$, with the theoretical spectrum in dashed lines. The two rectangles show the data range used in (b) and (c), with the ticks on the bottom indicating the $\omega_m/2\pi$ cuts. (b-c) Transmission amplitude cuts for different $\omega_m/2\pi$, with the spectrum in dashed lines. As in \cref{fig:cavity-pi:measurement:cuts}, the average value of the transmission is used to define the horizontal line from which the colouring begins. The fit parameters are indicated in \cref{tab:cavity-pi0} with the cavity resonances $\omega_{c,1}/2\pi = 6.594$ GHz, $\omega_{c,2}/2\pi=7.562$ GHz and $\omega_{c,3}/2\pi= 8.619$ GHz. The legend is common to all three subfigures and the curves are all offset by 50 dB.}
		\label{fig:cavity-pi0:measurement}
	\end{figure}
	\twocolumngrid\noindent
	obtained is that described in \cref{sec:coupling-phase}, and using \cref{fig:phase-reduction} we deduce that the physics is characterised by two coupling phases $\theta_1$ and $\theta_2$. Using \cref{tab:cavity-pi0}, we predict $\theta_1 = \varphi_{12}-\varphi_{11}-\qty(\varphi_{22}-\varphi_{21}) =\pi$ and $\theta_2 = \varphi_{32}-\varphi_{31}-\qty(\varphi_{22}-\varphi_{21}) =0$, hence the ``$\pi 0$'' cavity. The frequency-domain simulation is plotted in \cref{app:simulations} and agrees with the eigenmode analysis of \cref{tab:cavity-pi0}.

	\paragraph{Experimental results}
	For cavity $\pi$ of \cref{sec:single-phase}, we did not have a precise qualitative agreement with the theoretical predictions due to the additive manufacturing process of the cavity. Thus, we adopted a different strategy for cavity $\pi 0$: we used {brass} cylinders with a calibrated height to implement the posts, and we machined the rest of the cavity from {brass}. In re-entrant cavities, the eigenmodes are very sensitive to the distance between the top of the posts and the lid. Hence, by using posts with identical heights, we ensure that the potential error in the distance between the posts and the lid is uniform.
	
	The experimental data, along with the polariton spectrum, is plotted in \cref{fig:cavity-pi0:measurement}. This time, the fit parameters exactly correspond to those given in \cref{tab:cavity-pi0}, which  indicates that the construction of the cavity matches very well with the COMSOL design. As for the cavity $\pi$, the physical phase $\theta_1=\pi$ suggests that the eigen-frequencies do not cross between the first two anti-crossings, which is indeed confirmed by \cref{fig:cavity-pi0:measurement:cuts-1}. On the other hand, having $\theta_2=0$ should lead to a crossing of the eigen-frequencies between the second and third anti-crossings \cite{2023Gardin}. 	However when $\theta_2=0$ one of the polaritons is a magnonic dark mode \cite{2015Zhang,2023Gardin}, which does not lead to a resonance in the transmission amplitude \cite{2022Nair}. Thus, as observed in  \cref{fig:cavity-pi0:measurement:cuts-2}, we merely observe a single resonance peak moving as $\omega_m/2\pi$ is varied. This is verified using the input-output formalism \cite{1985Gardiner} in \cref{app:input-output}.

	Compared with cavity $\pi$ we notice the appearance of higher-order magnon modes, which manifests as anti-crossings located away from the magnon resonance $\omega_m/2\pi$. The excitation of these higher-order magnonic modes can be explained by the fact that the sphere used in this experiment are bigger (radius of 0.5mm, similar to the radius of the left and right posts, see \cref{fig:cavity-pi0-eigenmodes}), and thus the cavity modes magnetic fields of \cref{fig:cavity-pi0-magnetic} are less uniform over the spheres. The presence of a higher-order magnon mode can be verified by placing a single YIG sphere in cavity $\pi 0$, see \cref{fig:cavity-pi0-single} in \cref{app:single-yig-sphere}. Finally, as shown in \cref{app:single-yig-sphere}, we also observe the $\sqrt 2$ enhancement of the frequency gap at the anticrossings due to the coherent coupling of the two distant spin ensembles. \added[id=3]{The measured linewidths of the cavity modes are $\kappa_1/2\pi=17 \pm 0.5$ MHz, $\kappa_2/2\pi = 42.5 \pm 0.5$ MHz and $\kappa_3/2\pi = 20 \pm 0.5$ MHz. For the YIG spheres, we found the linewidths $\kappa_m/2\pi = 5 \pm 0.5$ MHz.}

	\section{\label{sec:conclusion}Conclusion}
	To summarise our results, we have proposed to use three-posts re-entrant cavities to engineer the coupling phases, which leads to engineering synthetic gauge fields. The experimental data was shown to match with numerical predictions based on the theory developed in \cite{2023Gardin}. Depending on the value of the physical phase $\theta$ which parametrises the physics, we observe either a cavity-mediated magnon-magnon coupling ($\theta=\pi$), or dark mode physics ($\theta = 0$). These findings are relevant for indirect coupling applications \cite{2019LachanceQuirion,2016Lambert,2016Hyde}, dark mode memories \cite{2015Zhang}, and the creation of non-reciprocal behaviours \cite{2023Biehs,2022Clerk,2015Metelmann}. 
	\added[id=5]{While we focused on the the two cases $\theta\in \qty{0,\pi}$, intermediate values of $\theta$ can be engineered using additional posts, as theoretically demonstrated in \cref{app:arbitrary-phases}.}
	
	Concerning indirect coupling, we verified the $\sqrt 2$ enhancement of the coupling strength due to having two distant macroscopic spin ensembles coherently coupling with the cavity mode. In the case of $\theta=\pi$, the magnon-magnon coupling extends in the dispersive regime, where both magnons are strongly detuned from the cavity modes. As shown in \cite{2023Gardin}, this results from the constructive interference of the cavity-mediated coupling by both cavity modes (while for $\theta=0$, the cavity-mediated couplings interfere destructively). It was shown that the dispersive regime is advantageous for sensing applications based on magnons (i.e. magnetometry \cite{2021Crescinia} or dark matter detection \cite{2019Flower}) so these schemes could benefit from engineering the coupling phases.
	
	This flexibility in the cavity-mediated coupling can also be useful for coupling magnons with superconducting transmon qubits through the cavity modes, so far only demonstrated in rectangular cavities \cite{2015Tabuchi,2016Tabuchi,2017LachanceQuirion,2020LachanceQuirion}. Indeed, such qubits couple to the electric field of the cavity modes, which in re-entrant cavities is focused on top of the posts see \cref{fig:cavity-pi-eigenmode-1-electric,fig:cavity-pi-eigenmode-2-electric,fig:cavity-pi0-electric}. The existence of several cavity modes, which may or may not couple to the magnons, is thus interesting for developing quantum information processing.

	Finally, note that here the gauge invariant phase $\theta$ parametrising the synthetic gauge field is fixed by the coupling phases, which are themselves uniquely determined by the geometry of the cavity. Nonetheless, tunability could be achieved by modulating the frequencies of one or more magnons \cite{2022Clerk}, for instance by directly driving a YIG sphere using a loop generating a magnetic field parallel to the static magnetic field. Hence, in principle, re-entrant cavities can be used to engineer a loop-coupled system, and by tuning the gauge-invariant phases, either through cavity engineering or driving the YIG spheres, we could steer the system towards either cavity-mediated or dark mode memory applications on-demand.

	\begin{acknowledgments}
		We would like to thank Bernard Abiven, mechanic at IMT Atlantique, for building cavity $\pi 0$. We acknowledge financial support from Thales Australia and Thales Research and Technology. This work is part of the research program supported by the European Union through the European Regional Development Fund (ERDF), by the Ministry of Higher Education and Research, Brittany and Rennes Métropole, through the CPER SpaceTech DroneTech, by Brest Métropole, and the ANR projects ICARUS and MagFunc. Jeremy Bourhill is funded by the Australian Research Council Centre of Excellence for Engineered Quantum Systems, CE170100009 and the Centre of Excellence for Dark Matter Particle Physics, CE200100008. The scientific colour map \textit{oslo} \cite{2021Crameri} is used in this study to prevent visual distortion of the data and exclusion of readers with colourvision deficiencies \cite{2020Crameri}.
	\end{acknowledgments}

    \appendix
    
    \begin{figure}[t]
    	\centering
    	\includegraphics[width=\linewidth]{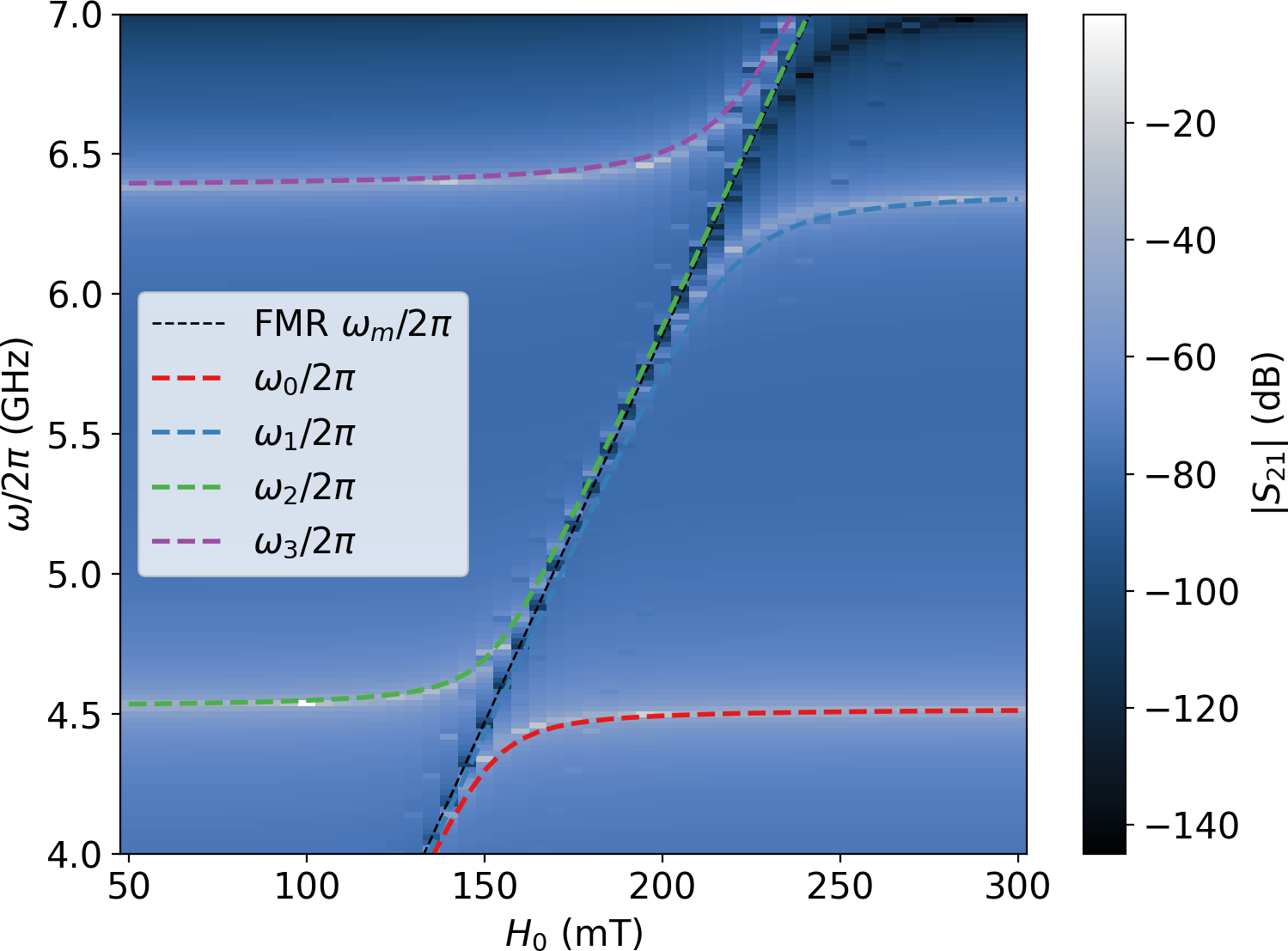}
    	\caption{Frequency-domain simulation using \comsol\ of the transmission through cavity $\pi$. The spectra predicted using the eigenmode analysis of COMSOL is plotted on top, with the eigen-frequencies given in \cref{fig:cavity-pi:all} and the other parameters given by \cref{tab:cavity-pi-comsol-theory}.}
    	\label{fig:simulation-fd:cavity-pi}
    \end{figure}
    
    \begin{figure}[t]
	    \centering
	    \includegraphics[width=\linewidth]{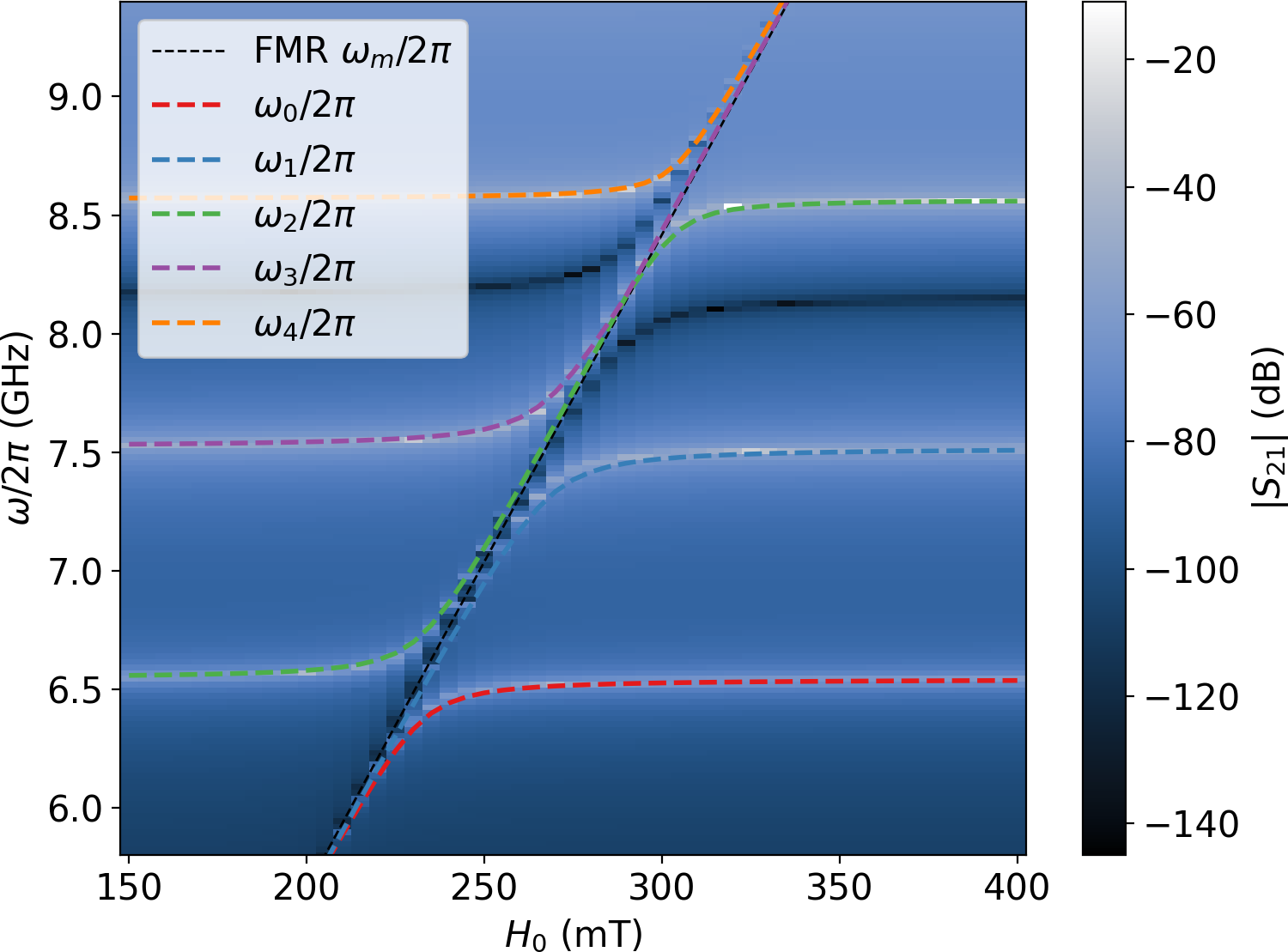}
	    \caption{Frequency-domain simulation using \comsol\ of the transmission through cavity $\pi 0$. The spectra predicted using the eigenmode analysis of COMSOL is plotted on top, with the eigen-frequencies given in \cref{sec:two-phases} of the main text, and other parameters given by \cref{tab:cavity-pi0}.}
	    \label{fig:simulation-fd:cavity-pi0}
    \end{figure}
    
    \section{\label{app:simulations}Simulation results}
    In this appendix, we compare the simulations results of the eigenmode and frequency-domain analyses of \comsol. The eigenmode simulations give the values of the coupling strengths and physical phases presented in \cref{tab:cavity-pi-comsol-theory,tab:cavity-pi0}, while the frequency-domain simulations give the transmission coefficient $S_{21}$. We use the results of the eigenmode analysis to plot the spectrum on top of the frequency-domain simulations, and we obtain good agreement, as shown in \cref{fig:simulation-fd:cavity-pi} for cavity $\pi$ and \cref{fig:simulation-fd:cavity-pi0} for cavity $\pi 0$.
    
    Importantly, the frequency-domain COMSOL simulations only include the Kittel mode for the magnons, and not higher-order modes. Therefore, the presence of crossings (anti-crossings) of the polaritonic frequencies for $\theta=0$ ($\theta=\pi$) observed cannot be due to the presence of higher-order modes.

    \section{\label{app:single-yig-sphere}Coherent coupling of two distant spheres}
\begin{figure}[b]
    	\centering
    	\includegraphics[width=\linewidth]{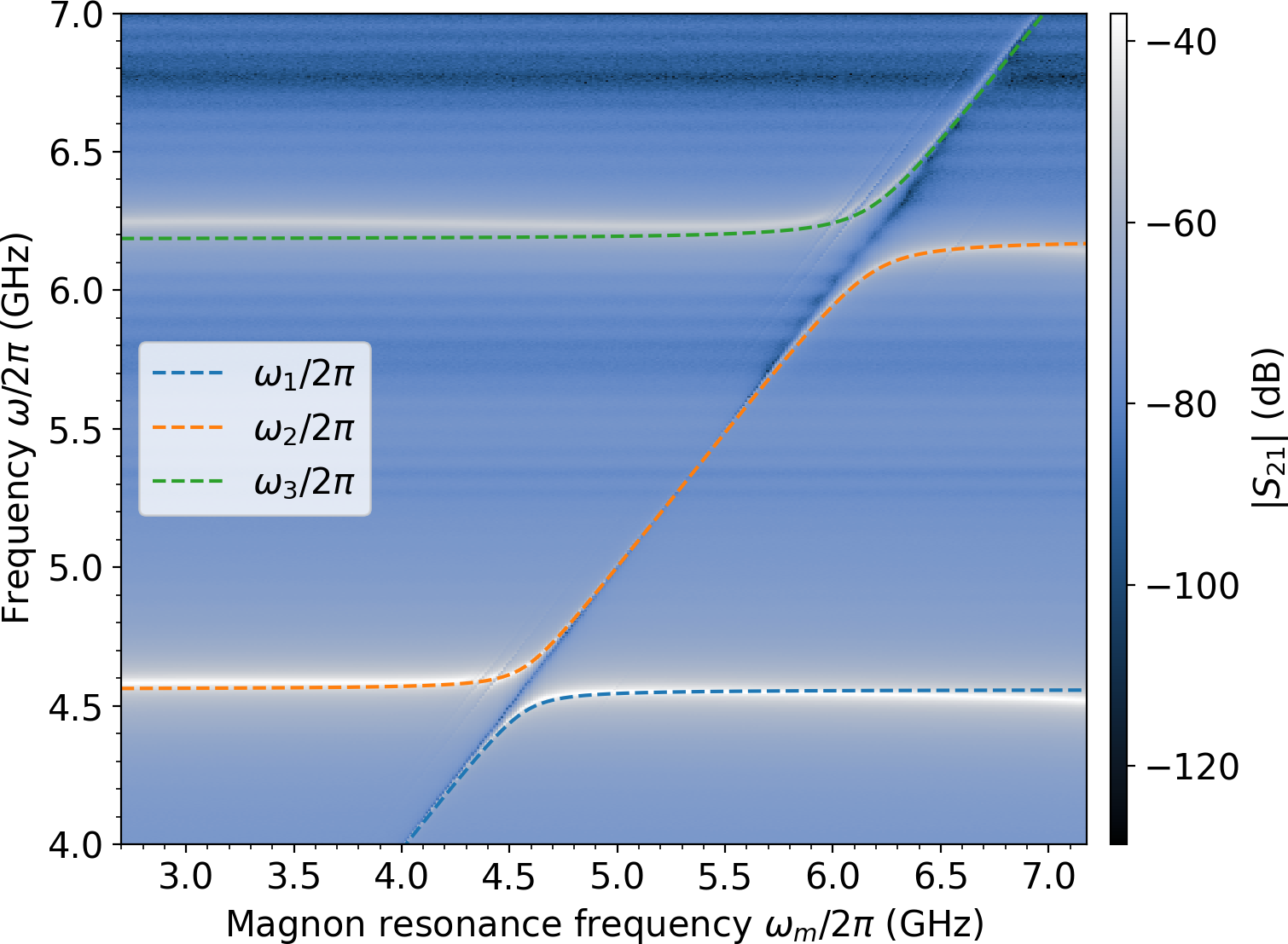}
    	\caption{Experimental transmission amplitude ($\abs{S_{21}}$) of cavity $\pi$ loaded with a single YIG sphere instead of two, with the theoretical spectrum in coloured lines. The fit parameters are identical to those of \cref{fig:cavity-pi:measurement:map}, except for a shift of the first cavity mode, set to $\omega_{c,1}/2\pi = 4.56$ GHz instead.}
    	\label{fig:cavity-pi-single}
    \end{figure}

\begin{figure}[t]
    	\centering
    	\includegraphics[width=\linewidth]{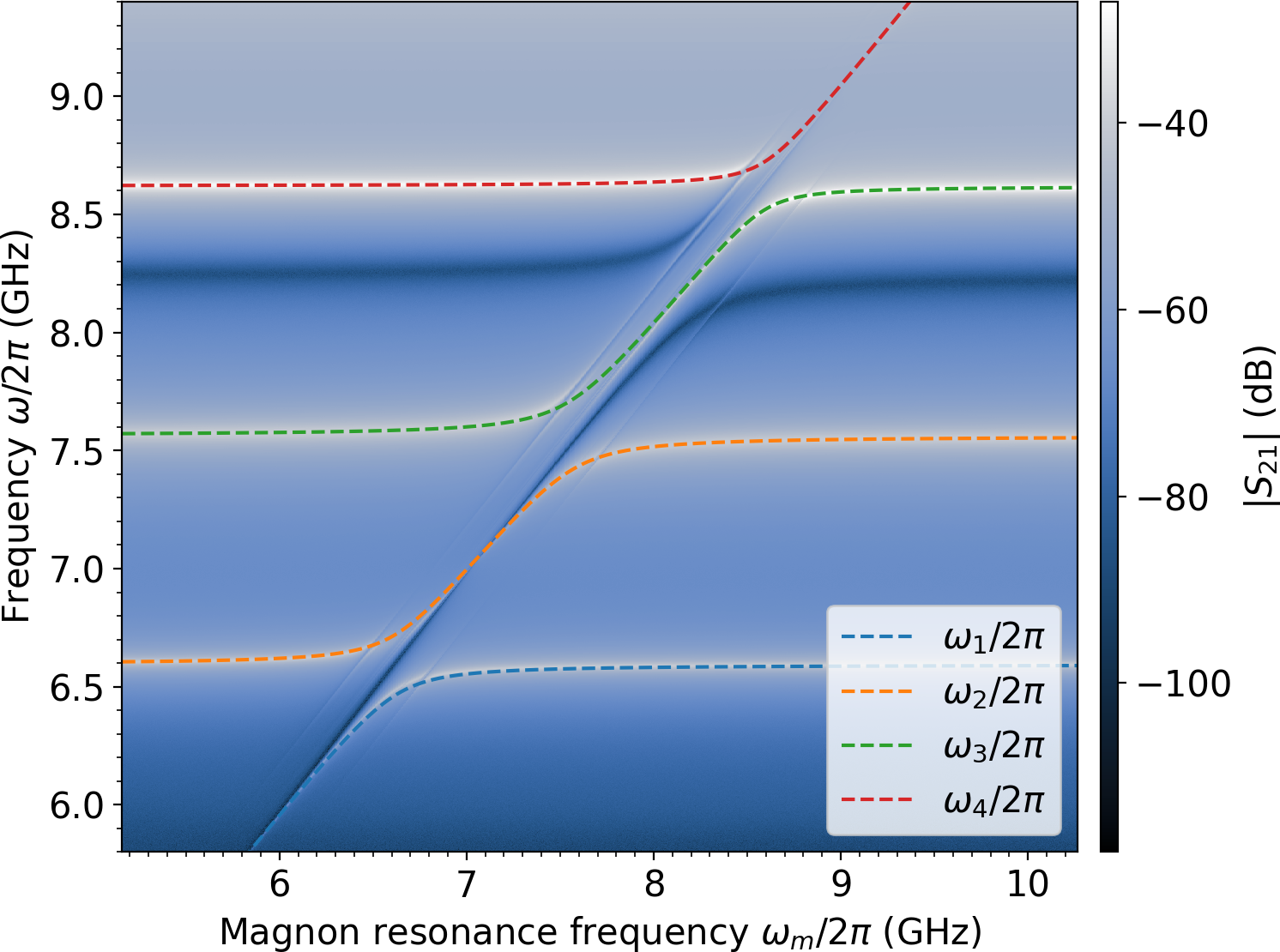}
    	\caption{Experimental transmission amplitude ($\abs{S_{21}}$) of cavity $\pi 0$ loaded with a single YIG sphere instead of two, with the theoretical spectrum in coloured lines. The fit parameters are exactly the same as those in \cref{fig:cavity-pi0:measurement}.}
    	\label{fig:cavity-pi0-single}
    \end{figure}

    The coherent coupling of $N$ YIG spheres to a cavity mode leads to an enhancement $\sqrt N$ of the frequency gap when the magnons are on resonance with the cavity mode \cite{2015Zhang}. To check that this is indeed the case, we removed one of the YIG spheres and measured again cavity $\pi$ and $\pi 0$. The experimental results are given in \cref{fig:cavity-pi-single} for cavity $\pi$ and \cref{fig:cavity-pi0-single} for cavity $\pi 0$. We can notice the presence of higher-order magnon modes in both figures, characterised by diagonal lines parallel to the Kittel mode's frequency $\omega/2\pi$.
    
\begin{figure}[t]
    	\centering
    	\includegraphics[width=\linewidth]{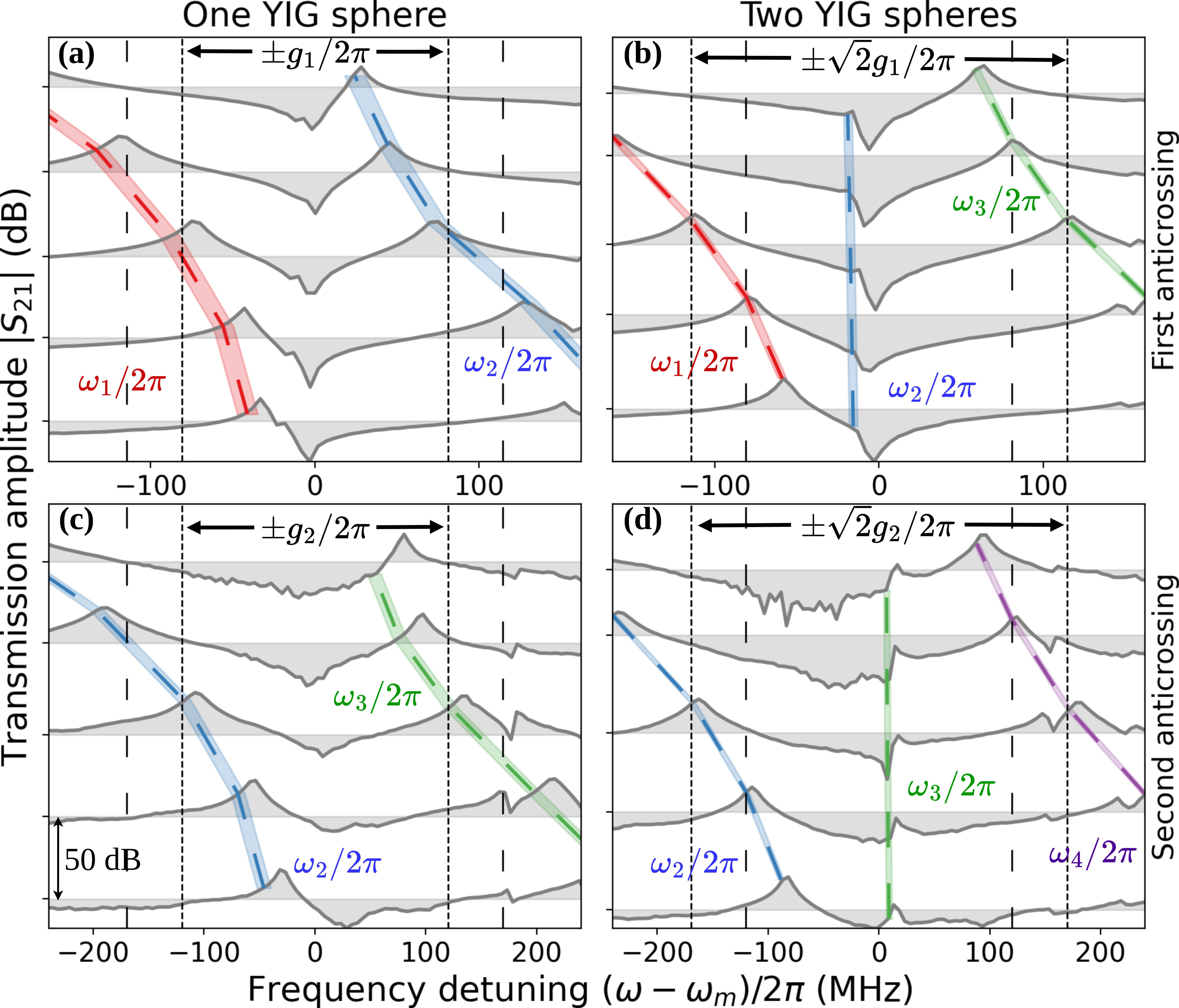}
    	\caption{Experimental transmission amplitude $\abs{S_{21}}$ of cavity $\pi$ for (a-b) the first anticrossing and (c-d) the second anticrossing. In the left column, we show the results for a single YIG sphere present in the cavity, while in the right column we show the case of two YIG spheres. From bottom to top, the line cuts in each subfigure are shown at $\omega_m \in \qty{\omega_{c,k}-2g_k, \omega_{c,k}-g_k, \omega_{c,k}, \omega_{c,k} + g_k, \omega_{c,k} + 2g_k}$. In all the subfigures, the curves are offset by 50 dB.}
    	\phantomsubfloat{\label{fig:cavity-pi-anticrossings:a}}
    	\phantomsubfloat{\label{fig:cavity-pi-anticrossings:b}}
    	\phantomsubfloat{\label{fig:cavity-pi-anticrossings:c}}
    	\phantomsubfloat{\label{fig:cavity-pi-anticrossings:d}}
    	\label{fig:cavity-pi-anticrossings}
    	\vspace{-4em}
    \end{figure}

    In \cref{fig:cavity-pi-anticrossings}, we compare the anti-crossings for cavity $\pi$ when a single YIG sphere is loaded in the cavity (\cref{fig:cavity-pi-anticrossings:a,fig:cavity-pi-anticrossings:c}), versus two YIG spheres  (\cref{fig:cavity-pi-anticrossings:b,fig:cavity-pi-anticrossings:d}). The dashed coloured lines track the polariton frequencies and are seen to follow the resonance peaks. In particular, the third line cut corresponds to the magnon(s) being on resonance with the cavity mode, for which the anticrossing frequency gap is minimum. We see that in the case of a single YIG sphere, this frequency gap is given by $2g_k/2\pi$ (first column), while for two YIG spheres the gap is $2\sqrt 2 g_k/2\pi$ (second column). 
    
\begin{figure}[t]
    	\centering
    	\includegraphics[width=\linewidth]{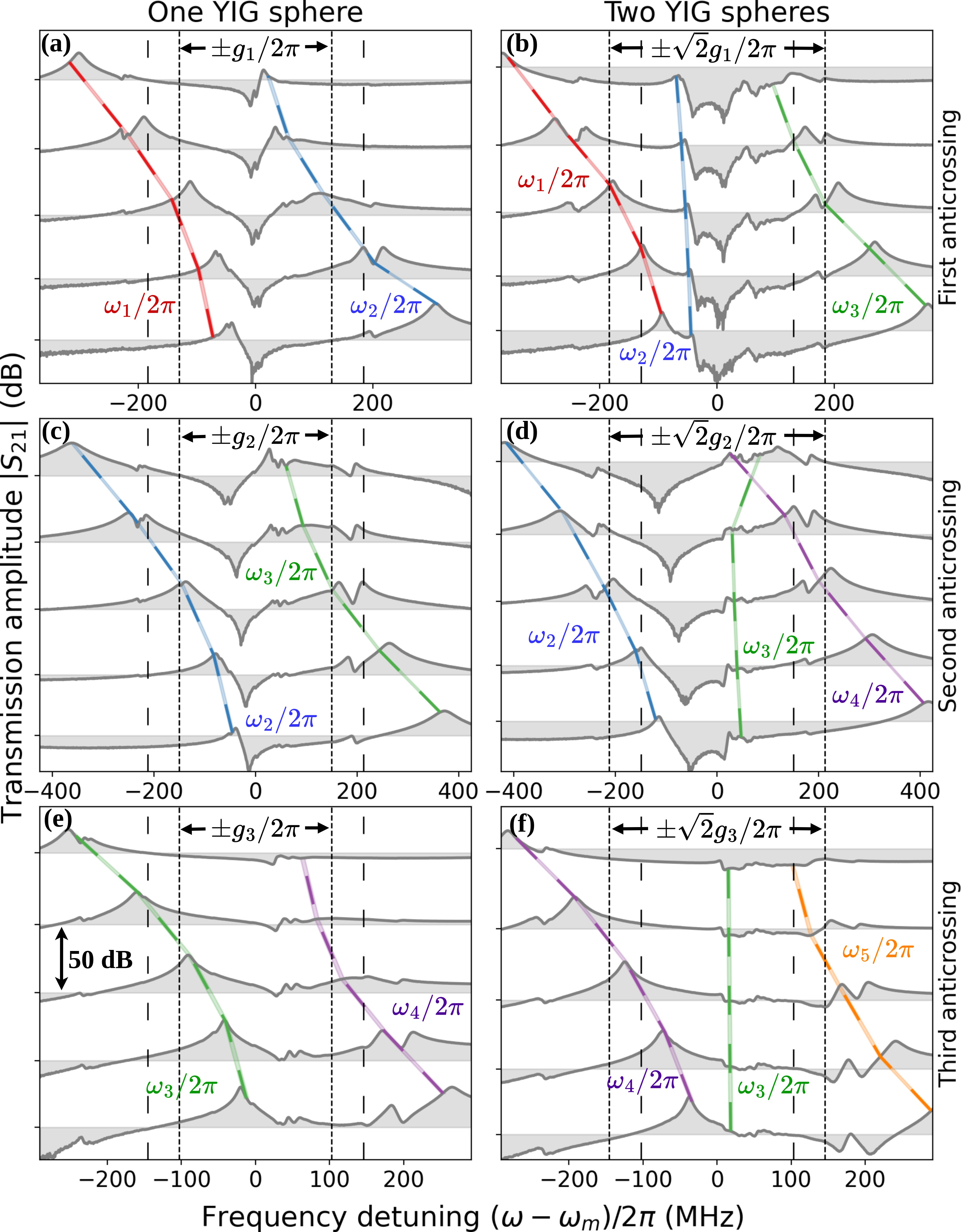}
    	\caption{Experimental transmission amplitude $\abs{S_{21}}$ of cavity $\pi 0$ for (a-b) the first anticrossing, (c-d) the second anticrossing and (e-f) the third anticrossing. The line cuts are identical to those of \cref{fig:cavity-pi-anticrossings}}
    	\phantomsubfloat{\label{fig:cavity-pi0-anticrossings:a}}
    	\phantomsubfloat{\label{fig:cavity-pi0-anticrossings:b}}
    	\phantomsubfloat{\label{fig:cavity-pi0-anticrossings:c}}
    	\phantomsubfloat{\label{fig:cavity-pi0-anticrossings:d}}
    	\phantomsubfloat{\label{fig:cavity-pi0-anticrossings:e}}
    	\phantomsubfloat{\label{fig:cavity-pi0-anticrossings:f}}
    	\label{fig:cavity-pi0-anticrossings}
    	\vspace{-3em}
    \end{figure}

    We carried out the same analysis for cavity $\pi 0$ in \cref{fig:cavity-pi0-anticrossings}, and the the $\sqrt 2$ enhancement is also verified. In \cref{fig:cavity-pi0-anticrossings:f}, the vertical lines indicating $\pm \sqrt 2 g_3$ seem to be offset to the left. Note that if a YIG sphere is removed, the system is not loop-coupled anymore, and thus the effects of the coupling phases disappear. 
    
    \section{\label{app:input-output}Input-output theory}
    The main objective of this appendix is to show that the dark mode arising for $\theta_2=0$ in cavity $\pi 0$ does not lead to a resonance in the transmission spectrum. In order to model the transmission amplitude $\abs{S_{21}}$ through the cavity, we use the input-output formalism \cite{1985Gardiner}, for which there is abundant literature, see \cite{2021HarderHu,2021Rameshti,2022Yuan} for instance. The derivation is rather lengthy \cite{2023Gardin}, and therefore we only describe the main steps here, and refer the reader to \cite{2024Bourcin} for more details.
    
    To model the intrinsic dissipation of the magnon modes in the YIG spheres, we couple each of them to a bosonic bath with a real-valued and frequency-independent coupling rate $\kappa_m/2\pi=5$ MHz (assumed to be identical for both YIG spheres). For the cavity modes, we only need to consider the two ports of the cavity, from which the S parameters are measured. Therefore, we assume that the cavity mode $c_i$ couples to two bosonic baths with frequency-independent coupling rates $\kappa_{ij} e^{i \phi_{ij}}/2\pi$, where $j \in \qty{1,2}$ indexes the bath. Note that the phases $\phi_{ij}$ are necessary to reproduce the anti-resonance behaviours, as recently shown by \cite{2024Bourcin}.
    
    \begin{figure}[t]
    	\centering
    	\includegraphics[width=\linewidth]{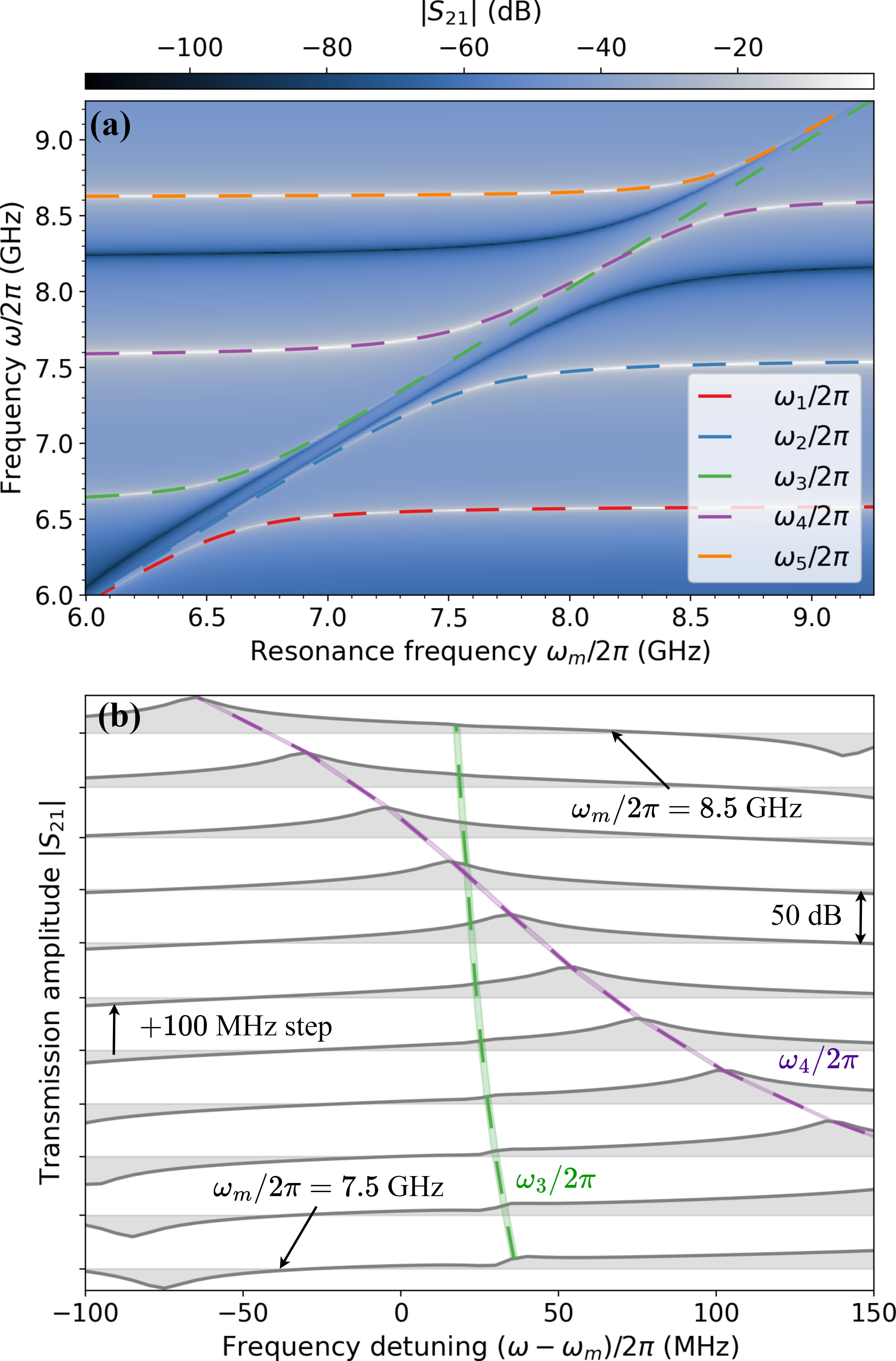}
    	\caption{Input-output modelling of the transmission amplitude $\abs{S_{21}}$ through cavity $\pi 0$. (a) Transmission amplitude reproducing the experimental results of \cref{fig:cavity-pi0:measurement:map}. (b) Cuts from $\omega_m/2\pi = 7.5$ GHz (bottom) up to $\omega_m/2\pi=8.5$ GHz, in steps of 100 MHz, of the transmission amplitude $\abs{S_{21}}$. This results mirrors the experimental data of \cref{fig:cavity-pi0:measurement:cuts-2} and showins the crossing of eigenvalues characteristic of $\theta_1=0$.}
    	\phantomsubfloat{\label{fig:inout-output:a}}
    	\phantomsubfloat{\label{fig:input-output:b}}
    	\label{fig:input-output}
    	\vspace{-3em}
    \end{figure}
    
    We use $\kappa_{1j}/2\pi = 1$ MHz, $\kappa_{2j}/2\pi=3$ MHz, and $\kappa_{3j}/2\i = 2$ MHz for $j \in \qty{1,2}$. The value of the phases $\phi_{ij}$ can be obtained by examining the direction of the magnetic field (for instance using an eigenmode analysis in COMSOL) coupled to the probes used to inject and measure the microwaves inside the cavity \cite{2024Bourcin}. We set $\phi_{1j}=\pi_{3j}=\pi$ for $j \in \qty{1,2}$n $\phi_{21}=\pi$ and $\phi_{22}=0$. Furthermore, the features of anti-resonances depend on several cavity modes, and to properly model the lowest frequency anti-resonance we included a lower frequency cavity mode with frequency $\omega_0/2\pi=3$ GHz, and the negative coupling rates $\kappa_{01}/2\pi = \kappa_{02}/2\pi = -5$ MHz to the ports. For the cavity resonance frequencies and coupling strengths, we use the same values as \cref{fig:cavity-pi0:measurement}. 
    
    The results are plotted in \cref{fig:input-output}, and in particular we note that \cref{fig:inout-output:a} reproduces successfully \cref{fig:cavity-pi0:measurement:map}. In \cref{fig:input-output:b}, we reproduce \cref{fig:cavity-pi0:measurement:cuts-2} in the absence of higher-
    \vspace{1em}
    \onecolumngrid
    
    \begin{figure}[hbt]
    	\includegraphics[width=\columnwidth]{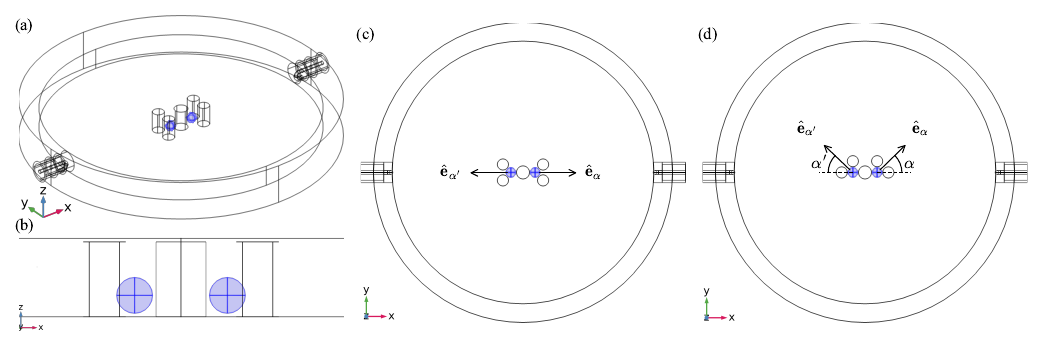}
    	\vspace{-1cm}
    	\caption{Representation of the {proposed }re-entrant cavity (a) in three dimensions, (b) in the ($z$, $x$) plane zoomed to view the gap $d$ between the top of the posts and the lid, (c) in the ($x$, $y$) plane with {the directions $\hat{\mathbf{e}}_\alpha,\hat{\mathbf{e}}_{\alpha'}$ represented where $\hat{\mathbf{e}}_{\alpha'}$ is the mirror symmetry of $\hat{\mathbf{e}}_\alpha$ with respect to the $(y,z)$ plane}, (d) in the ($x$, $y$) plane with the direction $\hat{\mathbf{e}}_\alpha$ making an angle $\alpha$ versus the $x$-axis. The two YIG sphere are represented in blue.}
    	\label{fig1}
    \end{figure}
    \twocolumngrid\noindent
    order magnon modes. As a consequence, the tracking of the polaritonic frequencies is simplified, and we verify the crossing of the eigenvalues. Notably, $\omega_3/2\pi$ does not lead to a resonance near the crossing, since it corresponds to a magnon dark mode as theoretically predicted in \cite{2023Gardin}.

	\section{\label{app:arbitrary-phases}\added[id=5]{Engineering arbitrary physical phases}}
	
	\begin{table}[t]
		\centering
		\caption{Characteristics of the first three cavity modes for $d = 50 µm$.}
		\begin{tabular}{c|ccccc}
			\label{tab1}%
			{Mode} & $\omega/2\pi$ (GHz) & ${\eta_\mathrm{L}}$ & ${\eta_\mathrm{R}}$ & ${g_\mathrm{L}/\omega}$ & ${g_\mathrm{R}/\omega}$ \\
			\hline
			0 & 4.02 & 0.018 & 0.019 & 0.011 & 0.012 \\
			1 & 8.29 & 0.125 & 0.123 & 0.055 & 0.056 \\
			2 & 12.03 & 0.144 & 0.145 & 0.055 & 0.055
		\end{tabular}
	\end{table}
	
	\added[id=5]{
	The objective of this investigation is to design a cavity capable of introducing a non-trivial phase between consecutive modes while operating in the strong coupling regime. To that effect, we developed a re-entrant cavity with five posts which additionally ensures uniform coupling strength across multiple YIG spheres. Furthermore, the phase relationship between the two modes is contingent upon the positioning of four posts within the cavity, as demonstrated in our analysis. By manipulating the positions of these posts, it is feasible to fine-tune the coupling phase by adjusting the gap between the top of the post and the cavity lid.}

	\added[id=5]{
	The cavity dimensions were optimized to maximize the coupling strength for each YIG sphere and to ensure that the phase falls within a range $\qty[n\frac{\pi}{4}, (n+1) \frac{\pi}{4}]$ with $n \in \mathbb{Z}$ by tuning an angle $\alpha$, as detailed below.
	As depicted in \cref{fig1}, the cavity is cylindrical and features five cylindrical posts. The cavity has a radius of 14 mm and a height of 2.25 mm. All posts have the same height, which depends on the value of the gap $d$; thus, the post height is $2.25 - d$. Positioned at the center of the cavity is a fixed post with a radius of 0.7 mm. The two YIG spheres are located on either side of the center post along the $x$-axis, with a spacing of 0.1 mm from the center post. Four additional posts are situated at the periphery of each YIG sphere, spaced 0.1 mm from the spheres. These peripheral posts are symmetrically positioned with respect to the $y$-axis, which passes through the center of the cavity. The radius of the peripheral posts is 0.6 mm. Relative to the direction $\hat{e}_\alpha$, the two peripheral posts on the right-hand side of the cavity are positioned at $\pm 45$°. Additionally, the direction $\hat{e}_\alpha$ forms an angle $\alpha$ with the $x$-axis, ranging from 0° to 45°, allowing the tuning of the coupling phases.
}

\begin{figure}[hbt]
		\includegraphics[width=\columnwidth]{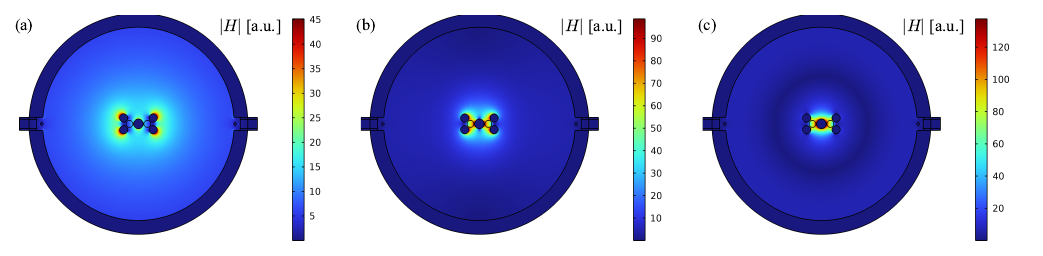}
		\vspace{-1cm}
		\caption{Representation of the $|H|$-field of the 3 first cavity modes in the ($x$, $y$) plane at the $z$-coordinate of the center of the YIG spheres.}
		\label{fig2}
	\end{figure}

	\added[id=5]{
	The norm of the cavity modes's magnetic fields are depicted in \cref{fig2} for the first three modes of the cavity. We focus on the second and third modes due to their high filling factor $\eta$ (see \cref{eq:filling-factor}), which are quite similar for both modes and spheres, as illustrated in \cref{tab1}. This table presents the eigenfrequencies $\omega/2\pi$ of the first three modes of the cavity, along with the filling factors $\eta_L$ and $\eta_R$ for the YIG spheres on the left and right-hand sides, respectively. Additionally, the associated coupling strengths $g_L$ and $g_R$ are provided. The values are given for an angle $\alpha = 0$° and a gap $d = 50$ µm.
}
	
	\added[id=5]{
	\Cref{tab2} presents similar characteristics for the two modes of interest, along with the averaged $H$-field direction inside each YIG sphere and the coupling phase. It is observed that the coupling strength slightly decreases from $\alpha = 0°$ to $\alpha = 45°$, while the difference in coupling strength between mode 1 and mode 2 slightly increases. However, the coupling strength remains relatively constant, ranging from 4.2\% to 5.7\%. The coupling phase varies between 129° and 180° for $\alpha$ ranging from 45° to 
	}
	
	\onecolumngrid
	
	\twocolumngrid\noindent
	\added[id=5]{0°, allowing for measurements with a non-trivial physical phase between 135° and 180°.}

	\begin{table}[h!]
		\centering
		\caption{Characteristics of the cavity modes versus $\alpha$ for $d = 50$ µm}
		\begin{tabular}{c|cccccccc}
			\label{tab2}%
			${\alpha}$ (°) & $\omega/2\pi$ (GHz) & ${\eta_\mathrm{L}}$ & ${\eta_\mathrm{R}}$ & ${g_\mathrm{L}/\omega}$ & ${g_\mathrm{R}/\omega}$ & ${\phi_\mathrm{L}}$ (°)& ${\phi_\mathrm{R}}$ (°) & ${\theta}$ (°) \\
			\hline
			\multirow{2}{*}{0} & 8.29 & 0.125 & 0.123 & 0.057 & 0.056 & 90 & 90 & \multirow{2}{*}{180} \\
			& 12.03 & 0.144 & 0.145 & 0.055 & 0.055 & -90 & 90 & \\
			\hline
			\multirow{2}{*}{5} & 8.29 & 0.125 & 0.122 & 0.057 & 0.056 & -91 & -88 & \multirow{2}{*}{170} \\
			& 12.09 & 0.141 & 0.142 & 0.053 & 0.053 & -86 & 87 & \\
			\hline
			\multirow{2}{*}{10} & 8.27 & 0.125 & 0.122 & 0.057 & 0.055 & -92 & -87 & \multirow{2}{*}{160} \\
			& 12.07 & 0.139 & 0.140 & 0.053 & 0.053 & -82 & 83 & \\
			\hline
			\multirow{2}{*}{15} & 8.33 & 0.124 & 0.122 & 0.056 & 0.055 & 86 & 95 & \multirow{2}{*}{151}\\
			& 12.09 & 0.134 & 0.134 & 0.050 & 0.051 & -79 & 80 & \\
			\hline
			\multirow{2}{*}{20} & 8.38 & 0.124 & 0.120 & 0.056 & 0.055 & 85 & 96 & \multirow{2}{*}{144} \\
			& 12.07 & 0.130 & 0.130 & 0.049 & 0.049 & -76 & 78 & \\
			\hline
			\multirow{2}{*}{25} & 8.48 & 0.122 & 0.121 & 0.055 & 0.054 & -97 & -83 & \multirow{2}{*}{139} \\
			& 12.05 & 0.126 & 0.125 & 0.047 & 0.047 & -76 & 77 & \\
			\hline
			\multirow{2}{*}{30} & 8.54 & 0.122 & 0.119 & 0.055 & 0.054 & 83 & 98 & \multirow{2}{*}{133} \\
			& 12.03 & 0.121 & 0.121 & 0.046 & 0.046 & -74 & 75 & \\
			\hline
			\multirow{2}{*}{35} & 8.62 & 0.121 & 0.119 & 0.054 & 0.053 & -98 & -81 & \multirow{2}{*}{131} \\
			& 11.99 & 0.119 & 0.119 & 0.045 & 0.045 & 106 & -105 & \\
			\hline
			\multirow{2}{*}{40} & 8.74 & 0.121 & 0.118 & 0.054 & 0.052 & 81 & 99 & \multirow{2}{*}{130} \\
			& 11.97 & 0.115 & 0.115 & 0.044 & 0.044 & -73 & 74 & \\
			\hline
			\multirow{2}{*}{45} & 8.90 & 0.120 & 0.118 & 0.053 & 0.052 & -99 & -80 & \multirow{2}{*}{128} \\
			& 11.96 & 0.112 & 0.112 & 0.042 & 0.042 & -74 & 74 & 
		\end{tabular}
	\end{table}

	\added[id=5]{
	\Cref{tab3} presents the same information for a fixed angle of 45° while varying the gap $d$ from 100 µm to 5 µm. It demonstrates that measurements with a non-trivial phase can be conducted within the range of 90° to 135°.
}
	\begin{table}[H]
		\centering
		\caption{Characteristics of the cavity modes versus $d$ for $\alpha = 45$°}
		\begin{tabular}{c|cccccccc}
			\label{tab3}%
			${d}$ (µm) & $\omega/2\pi$ (GHz) & ${\eta_\mathrm{L}}$ & ${\eta_\mathrm{R}}$ & ${g_\mathrm{L}/\omega}$ & ${g_\mathrm{R}/\omega}$ & ${\phi_\mathrm{L}}$ (°) & ${\phi_\mathrm{R}}$ (°) & ${\theta}$ (°) \\
			\hline
			\multirow{2}{*}{5} & 3.34 & 0.123 & 0.116 & 0.088 & 0.083 & 80 & 102 & \multirow{2}{*}{146}\\
			& 4.67 & 0.154 & 0.157 & 0.093 & 0.095 & 97 & -95 & \\
			\hline
			\multirow{2}{*}{10} & 4.63 & 0.124 & 0.115 & 0.075 & 0.070 & -100 & -78 & \multirow{2}{*}{147} \\
			& 6.45 & 0.155 & 0.158 & 0.080 & 0.082 & 97 & -94 & \\
			\hline
			\multirow{2}{*}{25} & 6.88 & 0.122 & 0.118 & 0.061 & 0.059 & -100 & -79 & \multirow{2}{*}{144} \\
			& 9.61 & 0.156 & 0.158 & 0.066 & 0.067 & 98 & -97 & \\
			\hline
			\multirow{2}{*}{50} & 8.90 & 0.120 & 0.118 & 0.053 & 0.052 & -99 & -80 & \multirow{2}{*}{129} \\
			& 11.95 & 0.112 & 0.112 & 0.043 & 0.043 & -74 & 74 & \\
			\hline
			\multirow{2}{*}{100} & 10.84 & 0.112 & 0.110 & 0.045 & 0.044 & 82 & 98 & \multirow{2}{*}{90}\\
			& 12.77 & 0.051 & 0.051 & 0.019 & 0.019 & -53 & 53 & 
		\end{tabular}
	\end{table}

	\bibliography{coupling-phase-cavity}
	
\end{document}